\documentclass[prl,showpacs,floatfix,twocolumn,superscriptaddress]{revtex4}

\usepackage{graphicx}
\usepackage{amsmath}
\usepackage{amssymb}
\usepackage{bm}

\begin{document}

\title{Dissipation of radial oscillations in compact stars}

\author{Basil A.\ Sa'd}
\email{basil.sad@gmail.com}
\affiliation{%
Institut f\"{u}r Theoretische Physik, Ruprechts-Karl-Universit\"{a}t, Philosophenweg 16, D-69120 Heidelberg, Germany}%
\affiliation{%
ExtreMe Matter Institute EMMI, GSI Helmholtzzentrum f\"{u}r Schwerionenforschung Gmbh, Planckstra{\ss}e 1, D-64291 Darmstadt, Germany.
}%
\author{J\"{u}rgen Schaffner-Bielich}
\email{Schaffner-Bielich@uni-heidelberg.de}
\affiliation{%
Institut f\"{u}r Theoretische Physik, Ruprechts-Karl-Universit\"{a}t, Philosophenweg 16, D-69120 Heidelberg, Germany}%
\affiliation{%
ExtreMe Matter Institute EMMI, GSI Helmholtzzentrum f\"{u}r Schwerionenforschung Gmbh, Planckstra{\ss}e 1, D-64291 Darmstadt, Germany.
}%

\date{\today}

\begin{abstract}
We demonstrate that there exists a new mechanism for dissipating the energy of stellar oscillations. For neutron stars, in particular, we show that the mechanical energy of density perturbations is not only dissipated to heat via bulk viscosity, but also that energy is radiated away via neutrinos. This energy dissipation will be associated with a viscosity coefficient, the radiative viscosity, which is larger than the bulk viscosity in the case of non-strange quark matter and nuclear matter.
\end{abstract}



\maketitle

{\it Introduction ---}Dissipative processes play an important role in the evolution of neutron stars. These processes are governed by transport coefficients, such as the heat and electrical conductivities, and the shear and bulk viscosities. The conductivities are important for stellar cooling as well as for the magnetic field decay. Shear viscosity damps differential rotation in a star and, thus, leads to a uniform rigid-body rotation. Bulk viscosity, on the other hand, damps the density oscillations of the stellar matter. Both differential rotation and oscillations could be excited in newly formed (hot) neutron stars, or could develop in old (cold) stars due to external perturbations, e.g., such as matter accreted from a companion star.


The viscosity of nuclear and mixed phases of dense baryonic matter has been calculated under various conditions and assumptions over the last three decades \cite{FlowersItoh1,FlowersItoh2,Sawyer,Jones1,Lindblom1,Lindblom2,Drago1,Haensel1,Haensel2,Chat}. The bulk viscosity of normal conducting strange quark matter was also calculated \cite{Sawyer2,Madsen,SSR2}. The latter might be relevant if the baryon density in the central regions of neutron stars is so high that matter becomes deconfined. 

 The main goal of this letter is to re-investigate the dissipation of density oscillations of high-density matter present in the interior of neutron stars. Using two-flavor quark matter as an example, it will be shown that the density perturbations are not only dissipated to heat via bulk viscosity, but also that energy is radiated away via neutrinos. This energy dissipation will be associated with a viscosity coefficient, the radiative viscosity (RV). We will also demonstrate that this RV coefficient is larger than the one for bulk viscosity.  

 We start by studying the effect of density oscillations on the composition of the system, mainly defined by the electron fraction. From that we will find that since the electron fraction oscillates out-of-phase with the baryon density, so will the pressure, and so the integrated $PdV$ work will result in a finite quantity which is the amount of energy of the oscillation converted to heat. This energy dissipation is associated with the bulk viscosity. Using the same formalism as for the bulk viscosity the increase of the energy radiated by neutrinos will be related to a viscosity coefficient which we call the radiative viscosity which we find to be closely related to the bulk viscosity for the type of interactions considered here for the \emph{urca} processes.

{\it Description of Oscillations ---}Let us begin by introducing a periodic perturbation to the baryon density $n(t)=n_0+{\mbox Re}\left(\delta n e^{i \omega t} \right)$, which will lead to a deviation from $\beta$-equilibrium characterized by $\delta \mu=\mu_{d}-\mu_{u}-\mu_e$

The density oscillations drive quark matter slightly out of $\beta$ equilibrium, regulated by the \emph{urca} processes;
\begin {equation}
d \to u+e+ \bar\nu_e,~~~ u+e \to d + \nu_e \label{interactions},
\end {equation}
\noindent but not out of thermal equilibrium which is restored almost immediately by strong processes. The corresponding instantaneous quasi-equilibrium state can be unambiguously characterized by the total baryon number density $n$ and the lepton fraction $X_e$,

\begin{equation}\label{composiiton}
n = \left(n_u+n_d\right)/3,~~~ X_e = n_e/n,
\end{equation}
\noindent where $n_u$ and $n_d$ are the number densities of up and down quarks, while $n_e$ is the number density of electrons. (In the case of strange quark matter, one should also add the strangeness fraction $X_s=n_s/n$ where $n_s$ is the number density of strange quarks \cite{SSR2}.) 
Charge neutrality requires
\begin{equation}
\frac23 n_u-\frac13 n_d-n_e=0. 
\label{neutral}
\end{equation}
Using this constraint together with the definitions in Eq.~(\ref{composiiton}), one can express the number densities and, in fact, all thermodynamic quantities of quark matter in terms of $n$ and $X_e$. For the number densities, for example, one finds
\begin{equation}
n_e = X_e n,~~ n_u =(1+X_e) n,~~ n_d =(2-X_e) n. 
\end{equation}
These number densities can also be expressed in terms of the corresponding chemical potentials, $n_i=n_i(\mu_i)$. In $\beta$ equilibrium, the three chemical potentials are related as follows: $\mu_d=\mu_u+\mu_e$. In pulsating matter, on the other hand, the instantaneous departure from equilibrium is described by the small parameter
\begin{equation}
\delta\mu \equiv \mu_d-\mu_u-\mu_e =  \delta\mu_d-\delta\mu_u-\delta\mu_e,
\label{delta-mu-def}
\end{equation}
where $\delta\mu_{i}$ denotes the deviation of chemical potential $\mu_i$ from its value in $\beta$ equilibrium. The quantity $\delta\mu$ can be conveniently expressed in terms of the variations of the two independent variables $\delta n$ and $\delta X_e$, 
\begin{equation}
\delta\mu = C\frac{\delta n}{n} + B \delta X_e,
\label{delta-mu}
\end{equation}
where, as follows from the definition, the coefficient functions $C$ and $B $ are given by 
\begin{subequations}
\begin{eqnarray}
C &=& n_d \frac{\partial \mu_d}{\partial n_d}
     -n_u \frac{\partial \mu_u}{\partial n_u}
     -n_e \frac{\partial \mu_e}{\partial n_e},   \label{def-C}\\
B  &=& -n \left(\frac{\partial \mu_d}{\partial n_d} 
                 + \frac{\partial \mu_u}{\partial n_u}
                 + \frac{\partial \mu_e}{\partial n_e} \right).
\label{def-B_e}
\end{eqnarray}
\label{def-C-B_e}
\end{subequations}
When $\delta\mu$ is non-zero the two \emph{urca} processes, c.f. Eq.~(\ref{interactions}), have slightly different rates. To leading order in $\delta\mu$, we could write
\begin{equation}
\Gamma_{\nu} - \Gamma_{\bar\nu} = - {\lambda} \delta\mu .
\label{ratediff0}
\end{equation}
 The net effect of having different rates for the two processes is a change of the electron fraction in the system:
\begin{equation}
n \frac{d (\delta X_e)}{dt} = {\lambda} \delta\mu ,
\label{ratediff}
\end{equation}
This has the tendency to restore the equilibrium value of $X_{e}$. Since the rate is finite, however, the weak processes always lag behind the density oscillations. In order to see this explicitly, we substitute $\delta\mu$ from Eq.~(\ref{delta-mu}) into Eq.~(\ref{ratediff}) and get the equation for $\delta X_e$ in a closed form,
\begin{equation}
n \frac{d (\delta X_e)}{dt} = {\lambda} 
\left(C\frac{\delta n}{n} + B \delta X_e\right) .
\label{eqX_e}
\end{equation}
The periodic solution to this equation can be found most easily by making use of complex variables. Denoting $\delta X_e \equiv \mbox{Re} \left(\delta X_{e,0}\, e^{i\omega t}\right)$, we derive the following result:
\begin{equation}
\delta X_{e,0} = \frac{\delta n_0}{n}\frac{C}{i\left( \frac{n \omega}{\lambda}\right)-B },
\label{deltaX_e0}
\end{equation}
 In the last equation, the lag of the weak processes is indicated by a non-vanishing imaginary part of $\delta X_{e,0}$. Such an imaginary part controls the phase shift of the $\delta X_e$ oscillations with respect to the oscillations of density. 

{\it Bulk viscosity ---}Let us now review the derivation of the bulk viscosity caused by weak interactions. This has been done for several possible phases of nuclear matter, see, for example, Refs \cite{Sawyer,SSR1,SSR2,Haensel1,Haensel2,Madsen,Alford:2006gy}.

For a periodic process, the bulk viscosity $\zeta$ is defined as the coefficient in the expression for the energy-density dissipation averaged over one period, $\tau=2 \pi/\omega$, 
\begin {equation}
\langle \dot{\cal E}_{\rm diss}\rangle =-\frac{\zeta}{\tau} 
\int_0^{\tau} dt \left(\nabla \cdot \vec v\right)^2,
\label{epsilon-kin}
\end{equation}
where $\vec v$ is the hydrodynamic velocity associated with the density oscillations. By making use of the continuity equation, $\dot{n}+n\,\nabla\cdot\vec v=0$, we derive
\begin{equation}
\langle \dot{\cal E}_{\rm diss}\rangle 
=-\frac{\zeta \omega^2}{2}\left(\frac {\delta n_0}{n}\right)^{2}.
\label{zeta-def}
\end{equation}
In order to solve for $\zeta$, the dissipated energy on the left-hand side has to be calculated explicitly. For the bulk viscosity, the energy dissipation is taken as the $PdV$ work done on the system.
\begin{equation}
\langle \dot{\cal E}_{\rm diss}\rangle 
= \frac{n}{\tau} \int_0^{\tau} P \dot{V} dt
\label{diss-energy}
\end{equation}
where $V\equiv 1/n$ is the specific volume. 

The pressure oscillations around the equilibrium value are driven by the
oscillations of its two independent variables, i.e., the quark number 
density and the lepton fraction,  
\begin{equation}
\delta P = \frac{\partial P}{\partial n}\, \delta n
-n\, C\, \delta X_e ,
\label{press}
\end{equation}
where $C$ is the same as in Eq.~(\ref{def-C}). In the derivation we took 
into account that $n_i =\partial P/\partial \mu_i$ and that the total 
pressure is given by the sum of the partial contributions of the quarks 
and electrons, $P=\sum_{i}P_{i}(\mu_i)$. 

Taking into account the relation (\ref{press}) together with the 
solution for $\delta X_{e,0}$ in Eq.~(\ref{deltaX_e0}), the expression
(\ref{diss-energy}) becomes
\begin{equation}
\langle \dot{\cal E}_{\rm diss}\rangle = -\frac{1}{2}\left(\frac{\delta n_0}{n}\right)^2\frac{\lambda\, \omega^2 C^2}
{\omega^2+\left(\lambda B /n\right)^2}.
\label{diss}
\end{equation}
By comparing this with the definition in Eq.~(\ref{zeta-def}), we finally derive 
an explicit expression for the bulk viscosity,
\begin{equation}
\zeta = \frac{\lambda C^2}{\omega^2+\left(\lambda B /n\right)^2} .
\end{equation}
For more details on the formalism and calculation of the bulk viscosity, the reader is advised to consult Refs.~ \cite{Sawyer,SSR1,SSR2,Haensel1,Haensel2,Madsen,Alford:2006gy}. One point worth mentioning here, is that this bulk viscosity coefficient controls the conversion of the mechanical energy of the oscillation into heat (recall that energy dissipation is given in terms of a $PdV$ work).

{\it Radiative Viscosity ---}Another source of dissipation, which has never been investigated before, is the change of neutrino emissivity due to the deviation from equilibrium. This energy dissipation will {\em not} increase the temperature, instead, the energy is radiated away in the form of neutrinos. We will associate a viscosity coefficient with this energy dissipation and refer to it as the radiative viscosity (RV).

 Since the change of interaction rates always leads to an increase of the neutrino emissivity, we can expand the off-equilibrium emissivity in terms of $\delta \mu^2$. To lowest order,
\begin{equation}
\dot E-\dot E_0 \equiv \delta \dot E ={\cal S}\delta \mu^2,
\end{equation}
where $\dot E_0$ is the neutrino emissivity at equilibrium as calculated in Refs.~\cite{Iwamoto,Iwamoto2}. 
 Using Eqs.~(\ref{deltaX_e0}) and (\ref{delta-mu}) we find, using $\delta \mu = \mbox{Re}\left(\delta \mu_0 e^{i \omega t}\right)$, that the volume oscillations are accompanied by an out-of phase oscillation of $\delta \mu$.

 The time-averaged change of neutrino emissivity is, then,
\begin{equation}
\langle \delta \dot E \rangle = \frac{\int_0^\tau {\cal S}(\delta \mu)^2 dt}{\tau} = \frac{\frac{{\cal S } C^2}{2}   \left( \frac{n \omega}{\lambda}\right)^2}{\left( \frac{n \omega}{\lambda}\right)^2+B^2} \left(\frac{\delta n_0}{n}\right)^2. 
\end{equation}
 This change of emissivity takes energy away from the oscillation and increases the rate of energy loss via neutrinos {\it i.e.} $\langle \dot {\cal E}_{diss}\rangle =-\langle \delta \dot E \rangle$.
 Since this energy loss is very similar to the energy loss due to the bulk viscosity, it is justifiable to define the RV coefficient (${\cal R}$) associated with this dissipation by using the same equation for the bulk viscosity c.f. Eq.~(\ref{zeta-def});
\begin{equation}
  {\cal R} = 2 \langle \delta \dot E  \rangle \left( \frac{\delta n_0}{n}  \right)^{-2} \omega^{-2}= \left( \frac{{\cal S}}{\lambda}  \right) \zeta.
 \end{equation} 
The dimensionless ratio ${\cal S}/\lambda$ can be found by examining the calculations for both the emissivity and the interaction rate. 
 Now let us turn to the calculation of $\lambda$ defined by 
Eq.~(\ref{ratediff0}). Following the original approach of Iwamoto 
\cite{Iwamoto}, we get the rate for $\beta$ decay in the following
form:
\begin{widetext}
\begin{equation}
\Gamma_{\bar \nu}(\delta \mu)= 
6\int \frac{d^3 p_d d^3 p_u d^3 p_{\bar\nu} d^3 p_e}{(2\pi)^{8} E_d E_u E_{\bar\nu} E_e}
\vert M \vert ^2 \delta ^4 \left(P_d-P_u-P_e-P_{\bar \nu}\right)
f\left({E_d-\mu _d}\right)\left[ 1-f\left({E_u-\mu _u}\right)
\right] \left[ 1-f\left({E_e-\mu _e}\right)\right].
\label{Gamma}
\end{equation}
\end{widetext}
Here, $P_i$ and $p_i$ are the 4- and 3-momenta of the $i$th particle, 
respectively, 
and $f(E)\equiv 1/(e^{E/T}+1)$ is the Fermi distribution function.
The scattering amplitude squared is given by \cite{Iwamoto}
\begin{eqnarray}
\vert M \vert ^2 &=& 64 G_F^2 \cos^2\theta_C(P_d\cdot P_{\bar \nu})(P_u \cdot P_e)\nonumber\\
&\approx&  \frac{2^8\alpha _s}{3 \pi} G_F^2 \cos^2\theta_C E_u E_d E_{\bar\nu} E_e 
(1-\cos\theta_{d\bar\nu}).
\end{eqnarray}
After substituting this approximate form of $\vert M \vert ^2$, all 
angular integrals in Eq.~(\ref{Gamma}) can be done exactly. Then, using 
the dimensionless variables $x_i=({E_i - \mu_i})/{T}$ (note that 
$\mu_{\bar\nu}=0$), we obtain
\begin{equation}
\Gamma_{\bar\nu}(\xi)=\frac{4\alpha _s}{\pi^6} G_F^2\cos^2 \theta_C 
\mu_d \mu_u \mu_e T^5  
\int_0^{\infty}dx_{\bar\nu}x_{\bar\nu}^2 J(x_{\bar\nu}-\xi),
\end{equation}
where $\xi \equiv {\delta \mu}/{T}$, and
\begin{equation}
J(x)=\left[ \prod_{j=1}^{3} \int_{-\infty}^{\infty}dx_j f(x_j)\right] 
\delta(x_1+x_2+x_3-x) = \frac{\pi^2+x^2}{2(1+e^x)}.
\end{equation}
By noting that 
$\Gamma_{\bar\nu}(\xi)=\Gamma_{\nu}(-\xi)$, we find 
\begin{equation}
\lambda = \frac{17}{15\pi^2} G_F^2\cos^2 \theta_C \alpha _s \mu_d \mu_u \mu_e T^4. 
\label{lambda}
\end{equation}

 In order to find the value of $\cal S$ we need to evaluate the off-equilibrium neutrino emissivity $\dot E=\dot E_{\bar \nu}+\dot E_\nu$ where $\dot E_{\bar \nu}$ and $\dot E_\nu$ are the energy loss rates via anti-neutrinos and neutrinos, respectively.
 The integral for the energy loss rate due to anti-neutrinos $\dot E_{\bar \nu}$ is the same as the one in Eq.~(\ref{Gamma}) except that the integrand is multiplied by the energy of the anti-neutrinos $p_{\bar \nu}$, and so, the angular integral and the integral over the momenta of quarks and electrons are the same;
\begin{equation}
\dot E_{\bar\nu}(\xi)=\frac{4\alpha _s}{\pi^6} G_F^2\cos^2 \theta_C 
\mu_d \mu_u \mu_e T^6  
\int_0^{\infty}dx_{\bar\nu}x_{\bar\nu}^3 J(x_{\bar\nu}-\xi).
\end{equation}
Noting that $\dot E_{\bar \nu}(\xi)=\dot E_{\nu}(-\xi)$, we derive
\begin{equation}
\dot{E} = \frac{4\alpha_s}{\pi^6}G_F^2 \cos^2(\theta_c)\mu_d \mu_u \mu_e T^6 \left[ \frac{457 \pi^6}{2520}+\frac{17 \pi^4}{40} \left( \frac{\delta \mu}{T}\right)^2  \right].
\label{emissivity}
\end{equation}
The first term in this equation is the equilibrium emissivity. We finally find that
\begin{equation}
\frac{{\cal S}}{\lambda}=\frac 3 2.
\label{ratio}
\end{equation} 

We stress that this ratio is generic to all \emph{urca} processes, both in nuclear matter and quark matter, see, for instance Refs.~\cite{NENS,TEPNS}.

{\it Discussion ---}In this letter we discussed the effect of \emph{urca} processes on the dissipation of density perturbations in the star. It was shown that they do not only contribute by converting energy into heat ``via bulk viscosity'', but also that they dissipate the energy by converting it into an increase of the neutrino emissivity. 

 We have associated this energy dissipation with a viscosity coefficient, the radiative viscosity (RV), which, except for not converting the energy of the perturbation into heat, behaves exactly like a bulk viscosity coefficient: the mechanical energy of the perturbation is radiated away from the star in the form of neutrinos.  We have shown that in the case of non-strange quark matter, this effect is $1.5$ times larger than the one of the bulk viscosity. This result also holds for the case of a proton-neutron-electron system.

 In the case of strange quark matter, the radiative dissipation is only governed by the \emph{urca} processes of both the strange quark and the down quark. The non-leptonic weak interactions will clearly not contribute to the radiative viscosity (see Ref.~\cite{SSR2} for a detailed analysis of the contribution of both the non-leptonic and \emph{urca} processes to the bulk viscosity) which may complicate the situation since the non-leptonic weak interactions dominate the bulk viscosity energy dissipation for a wide range of temperatures and oscillation frequencies \cite{SSR2}. 
 
 The radiative viscosity will contribute to the dissipation of density oscillations in the star, these oscillations may arise during phase transitions as in Ref.~\cite{Migdal}, or due to giant flares in magnetars (highly magnetized neutron stars), see Ref.~\cite{Watts}, or instabilities that result from the emission of gravitational waves \cite{Chandra1,Chandra2,Frid1,Andersson,Fried2}. The so-called r-mode (or rotation-dominated) instabilities might be the most important ones. They can develop at a relatively low angular velocity \cite{Lind1, Anderssonreview, Lind-lect, Madsenprl2}, and therefore may be relevant for a large number of compact stars. The presence of viscosities relieves this situation by damping these instabilities, prohibiting the star from emitting gravitational waves and loosing angular momentum.
 
 We note that the dissipation of radial perturbations via an increase of the energy loss to radiation is a general phenomenon applicable to all physical processes that involve transfer of energy to radiation. Such processes are relevant, for instance, for variable stars, which change their luminosity due to density oscillations, see Refs.~\cite{Cunha1, Cunha2}. Examples for such stars are pulsating white dwarfes or asymptotic giant branch (AGB) stars. This change of luminosity may be associated with a radiative viscosity coefficient responsible for dissipating these oscillations. In this case the energy is radiated away by photons from the surface rather than neutrinos. One has to be cautious, though, that the formalism used in this letter assumes averaging over one period of oscillation, which may give zero in the case of variable stars, so that the formalism might have to be changed to account for this fact. 

{\bf Acknowledgements} The authors acknowledge discussions with Giuseppe Pagliara, Giorgio Torrieri, Jorge Noronha, Igor Shovkovy, Matthias Hempel, Debarati Chatterjee, and Anna Watts. This work was supported by the Helmholtz Alliance Program of the Helmholtz Association, contract HA216/EMMI ``Extremes of Density and Temperature: Cosmic matter in the laboratory and BMBF grant FKZ 06HD9127''.
\bibliography{radial2}
\end{document}